\documentclass{aastex6}
\newcommand{\be}{\begin{equation}}
\newcommand{\ee}{\end{equation}}
\def\ergs{{\rm\,erg\,s^{-1}}}
\newcommand{\msun}{{M}_{\sun}}  

\fullcollaborationName{The Friends of AASTeX Collaboration}

\begin{document}


\title{The possible submillimeter bump and accretion-jet in the central supermassive black hole of NGC 4993}

%% Use \author, \affil, plus the \and command to format author and affiliation
%% information.  If done correctly the peer review system will be able to
%% automatically put the author and affiliation information from the manuscript
%% and save the corresponding author the trouble of entering it by hand.
%%
%% The \affil should be used to document primary affiliations and the
%% \altaffil should be used for secondary affiliations, titles, or email.

%% Authors with the same affiliation can be grouped in a single
\author{Qingwen Wu\altaffilmark{1,4}, Jianchao Feng\altaffilmark{2,3}, and Xuliang Fan\altaffilmark{1,3}}
%\and
\altaffiltext{1}{School of Physics, Huazhong University of Science and Technology, Wuhan, 430074, China}
\altaffiltext{2}{School of Physics and Electronic Science, Guizhou Normal University, No. 116, Baoshan Road, Guiyang 550001, China}
\altaffiltext{3}{Guizhou Provincial Key Laboratory of Radio Astronomy and Data Processing, Guizhou Normal University, Guiyang 550001, China}
\altaffiltext{4}{Corresponding author E-mail: qwwu@hust.edu.cn}

%\altaffiltext{5}{these authors contributed equally to this work}


\begin{abstract}
NGC 4993, as a host galaxy of the electromagnetic counterpart of the first gravitational-wave detection of a binary neutron-star merger, was observed by many powerful telescopes from radio to $\gamma$-ray waveband. The weak nuclear activities of NGC 4993 suggest that it is a low-luminosity active galactic nuclear (LLAGN). We build the multi-waveband spectral energy distributions (SEDs) of NGC 4993 from literatures. We find that the radio spectrum at $\sim 100-300$ GHz is much steeper than that of low-frequency waveband (e.g., 6-100 GHz), where this break was also found in the supermassive black holes in our galaxy center (Sgr A*), and in some other nearby AGNs. The radio emission above and below this break may has different physical origins, which provide an opportunity to probe the accretion and jet properties. We model the multi-waveband SEDs of NGC 4993 with an advection-dominated accretion flow (ADAF)-jet model. We find that the high-frequency steep radio emission at millimeter waveband is consistent with the prediction of the ADAF, while the low-frequency flat radio spectrum is better fitted by the jet. Further more, the X-ray emission can be also explained by the ADAF model simultaneously. From the model fits, we estimate important parameters of the central engine (e.g., accretion rate near the horizon of the black hole and mass-loss rate in jet) for NGC 4993. This result strengthen that the millimiter, submillimer and deep X-ray observations are crucial to understand the weak or quiescent activities in supermassive black-hole systems. Further simultaneous millimeter and X-ray monitoring of this kind of LLAGNs will help us to better understand the physical origin of multiwaveband emission.
\end{abstract}

\keywords{accretion, accretion disks - black hole physics - galaxies: jets - galaxies:individual (NGC 4993).}

\section{Introduction} \label{sec:intro}
It is now widely believed that most, if not all, galaxies harbor a supermassive black hole (SMBH) with a mass of $10^6-10^{10} \msun$. Different level of nuclear activities are found in different types of galaxies, ranging from most active and luminous active galactic nulei (AGNs), to less active low-luminosity AGNs (LLAGNs) and until the quiescent normal galaxies as our own. The activities of different types of galaxies mainly determined and/or co-evolved with the activities of the central SMBH. The most luminous AGNs (e.g., quasars and narrow-line Seyfert Is) host a SMBH accreting at sub-Eddington rate through the standard disk or even with a super Eddington rate through slim disk \citep[e.g.,][]{ss73,ab88,wang14}. For nearby LLAGNs or even quiescent normal galaxies (e.g., weak Seyfert, LINERs), the central SMBHs normally accreting through advection-dominated accretion flow (ADAF), where most of the gravitational energy released by accreting matter is advected into cenrtal BH when accretion rate is less than a critical value of $\sim$1 percent of Eddington rate \citep[e.g.,][]{ny95,ab95,yn14}. The strongly evolved accretion processes are also found in stellar-mass BH X-ray binaries (XRBs), where the X-ray spectra also strongly evolved during outburst \citep[e.g.,][]{wu08} and shown similar features as those in AGNs \citep[e.g.,][]{wang04,gu09}. 

Compact radio cores and/or weak jets are widely observed in nearby LLAGNs\citep[e.g.,][]{fm00,ho02}. The radio emission is normally much higher than the radiation of the thermal electrons ADAF and is normally more consistent with a jet/wind origin or emission from nonthermal electrons in ADAF \citep[e.g.,][]{yuan03,liu13}. This phenomenon is also found in XRBs, where the radio emission is stronger in low/hard state while it become weaker or disappear in high/soft state \citep[e.g.,][]{cob03}. Due to the appearance of the jet in LLAGNs and low/hard state XRBs, the origin of their multi-waveband emission is always in debate. \cite{mar05,mar08} proposed that most of radiation from radio to X-rays may be dominated by the jet in low accretion regime. However, both ADAF and jet contribution are considered in modeling the multi-waveband spectral energy distribution (SED) of LLAGNs \citep[e.g.,][]{wu07,yu11,nem14}, where the radio emission is always dominated by the jet while the X-ray emission is either dominated by the ADAF or by the jet depending on the Eddington ratio \citep[e.g.,][]{yc05}. The millimeter excess was found in both nearby LLAGNs and also in some bright Seyferts \citep[e.g.,][]{an05,di05,di11,di16,be15,pri16}. This excess suggests the possible different physical origins compared the low-frequency radio emission.
 \citet{di11} suggested that the millimeter emission should come from the AGN cores, where the other possibilities are also discussed, but is unlikely. The synchrotron emission from the hot thermal electrons in ADAF radiate at submm waveband, which was used to constrain the accretion flow in Sgr A* and M 87 \citep[e.g.,][]{yuan03,feng16}. \citet{be15} suggested that the high-frequency excess in millimeter waveband should be correlated to the accretion flow (e.g., corona produced by the magnetic activity around the accretion disk) based on 95 GHz observation on a sample of radio-quiet AGNs. \citet{ba15} found the millimeter variability is similar to that of X-ray emission in NGC 7469, which further support that they may have the same origin (e.g., associated with the accretion disk corona).

NGC 4993 is a nearby S0 galaxy, which is the host galaxy of the EM counterpart for the GW 170817A \citep[][]{ab17}. It has a redshift of $z=0.009873$, corresponding to a distance of $\simeq 40$ Mpc \citep[][]{lev17}. The weak [OIII], [NII], [SII] emission lines are presented in nucleus, and the relatively high ratio of $\rm [NII]\lambda6583/H\alpha$ is suggestive of a LLAGNs rather than star formation \citep[may be also driven by some hot post-AGB stars or shocks, e.g., ][]{lev17}. The nuclear activities are also clearly detected at other wavebands \citep[e.g., radio, submillimeter and X-rays][]{kim17,hag17}, which suggest that the central BH still show weak activities and not fully go into the quiescent state. 

The accretion-jet physics has been widely explored in LLAGNs, where the radio and X-ray wavebands are normally used to constrain the models \citep[e.g.,][]{wu07,yuan09,nem14}. The millimeter observation also play an important role in helping us to understand the BH activities. However, the millimeter observations, particularly high-resolution observations as ALMA, are still rare for BH sources. Fortunately, many ground and space telescopes observed and monitored the event of GW 170817A due to it is first electromagnetic counterpart of the gravitational-wave\citep[][]{ab17}. The observational data, including the millimeter observations by ALMA, on its host galaxy was also fruitful, which can help us to explore the SMBH activities. In this work, we explore the accretion-jet process for the SMBH in NGC 4993 based on the most recent multi-waveband observations, where the submm and deep $Chandra$ observations can help us to understand the accretion-jet properties for the SMBH in the weak or quiescent state.

\section{Data}
   The GW 170817A and its host galaxy were observed by the Very Large Array (VLA) and Atacama Large Millimeter/sub-millimeter Array (ALMA) from August 18 ($\sim$ half day after the GW event) to September 25. The radio emission of the host galaxy is unresolved in both VLA and ALMA observations at resolution of $\sim 0.1^{"}-1^{"}$ (corresponding to $<20-200$ pc at a distance of 40 Mpc). During above observations for $\sim$ one month, both VLA and ALMA observations show $\sim20\%$ variations, which suggest the activities of the central BH\citep[][]{ax17}. The imaging from Very Long Baseline Array (VLBA) also find an unresolved core or a marginally-resolved source on a scale smaller than the VLBA synthesized beam (2.5$\times$1.0 mas) with a 9$\sigma$ flux density of 0.22$\pm0.04$ mJy and a brightness temperature of  $1.6\times10^6$ K \citep[][]{de17}. \citet{hag17} presented two deep $Chandra$ observations on September 1-2, where both GW counterpart and its host galaxy are detected. The compact X-ray source is consistent with the nucleus of the galaxy with a hard X-ray spectrum(photon index $\Gamma=1.5\pm0.4$). This X-ray emission is most likely due to a weak LLAGN. We also include the $Hubble\ Space\ Telescope$ (HST), $2MASS$,  $Spitzer$, $Pan$-$STARRS$(PS1), $GALEX$, $WISE$ and $Very\ Large\ Telescope$ (VLT) observations, where the photometry has been meaured using Kron apertures or $1^{'}$ apertures centered on the host galaxy\citep[][]{lev17,ba17}. We list all the selected data for building the SED in Table (1).

\begin{table*}[]
\centering
  \caption{The Multi-wavelength data of NGC 4993.}
  \label{data}
  \begin{tabular}{cccccc}
  \hline
  Telescope & Band & Frequency & $\nu L_{\nu}$ (erg s$^{-1}$) & Aperture/Resolution (\arcsec) & Reference \\
  \hline
         VLA &     6 GHz  &  $6.00\times 10^{9}$  &  $(3.70\pm0.22)\times 10^{36}$ &     $\sim1$ & 1 \\
         VLA &   9.7 GHz  &  $9.70\times 10^{9}$   &  $(4.53\pm1.00)\times 10^{36}$ &     $\sim1$ & 1 \\
         VLA &    10 GHz  &  $1.00\times 10^{10}$  &  $(5.38\pm0.37)\times 10^{36}$ &     $\sim1$ & 1 \\
         VLA &    15 GHz  &  $1.50\times 10^{10}$  &  $(8.26\pm0.50)\times 10^{36}$ &     $\sim1$ & 1 \\
        VLBA &   8.7 GHz  & $8.70\times 10^{9}$   &  $(3.57\pm0.65)\times 10^{36}$ &     $\sim0.002$ & 2 \\
        ALMA &  97.5 GHz  &  $9.75\times 10^{10}$  &  $(3.82\pm0.36)\times 10^{37}$ &     $\sim0.2$ & 1 \\
        ALMA & 338.5 GHz  &  $3.39\times 10^{11}$  &  $(6.76\pm1.33)\times 10^{38}$ &     $\sim0.1$ & 3 \\
        WISE &  22$\mu$m  &  $1.36\times 10^{13}$  &  $(2.64\pm0.44)\times 10^{41}$ &     Kron aperture & 1 \\
        WISE &  12$\mu$m  &  $2.50\times 10^{13}$  &  $(5.61\pm0.21)\times 10^{41}$ &     Kron aperture & 1 \\
        WISE &  4.6$\mu$m &  $6.52\times 10^{13}$  &  $(4.07\pm0.03)\times 10^{42}$ &     Kron aperture& 1 \\
        WISE &  3.4$\mu$m &  $8.82\times 10^{13}$  &  $(1.02\pm0.01)\times 10^{43}$ &     Kron aperture & 1 \\
Spitzer/IRAC & 4.5 $\mu$m &  $6.66\times 10^{13}$  &  $(5.23\pm0.10)\times 10^{42}$ &      60$^a$ & 4 \\
Spitzer/IRAC & 3.6 $\mu$m &  $8.33\times 10^{13}$  &  $(1.07\pm0.02)\times 10^{43}$ &      60$^a$ & 4 \\     
       2MASS &  Ks        &  $1.38\times 10^{14}$  &  $(3.66\pm0.06)\times 10^{43}$ &     Kron aperture & 1 \\         
       2MASS &  H         &  $1.82\times 10^{14}$  &  $(5.79\pm0.11)\times 10^{43}$ &     Kron aperture & 1 \\    
       2MASS &  J         &  $2.40\times 10^{14}$  &  $(6.59\pm0.12)\times 10^{43}$ &     Kron aperture & 1 \\   
    HST/WFC3 &     F160W  &  $1.88\times 10^{14}$  &  $(5.36\pm0.05)\times 10^{43}$ &      60$^a$ & 4 \\
    HST/WFC3 &     F110W  &  $2.72\times 10^{14}$  &  $(5.72\pm0.05)\times 10^{43}$ &      60$^a$ & 4 \\
    HST/WFC3 &     F275W  &  $1.09\times 10^{15}$  &  $(8.33\pm1.15)\times 10^{40}$ &      60$^a$ & 4 \\
  VLT/HAWK-I &         K  &  $1.37\times 10^{14}$  &  $(2.33\pm0.02)\times 10^{43}$ &      60$^a$ & 4 \\
   VLT/VIMOS &         Z  &  $3.30\times 10^{14}$  &  $(5.12\pm0.05)\times 10^{43}$ &      60$^a$ & 4 \\
   VLT/VIMOS &         R  &  $4.68\times 10^{14}$  &  $(4.46\pm0.04)\times 10^{43}$ &      60$^a$ & 4 \\
   VLT/VIMOS &         U  &  $8.19\times 10^{14}$  &  $(5.76\pm0.05)\times 10^{42}$ &      60$^a$ & 4 \\
     Chandra &   0.3 keV  &  $7.25\times 10^{16}$  &  $(2.91\pm2.52)\times 10^{38}$ &      $\sim0.5$ & 5 \\
     Chandra &     8 keV  &  $1.93\times 10^{18}$  &  $(1.50\pm0.72)\times 10^{39}$ &      $\sim0.5$ & 5 \\
  \hline
  \end{tabular}
 
  \tablecomments{$^a$ The photometry has been measured in $60^{''}$ apertures centered at the host galaxy; \\
   References: 1): \citet{ba17}, 2): \citet{de17}, 3): \citet{kim17}, 4): \citet{lev17}, 5): \citet{hag17} }
\end{table*}

\section{ADAF and jet model}  
  Due to the weak activities in NGC 4993, the SMBH should accrete the surrounding material through the ADAF. We simply introduce the model as below, and more details can be found in \citet{yc05}, \citet{wu07} and \citet{feng16}. 
  
   We numerically solve the global structure of the ADAF, where the ion and electron temperature, density, angular momentum, radial velocity at each radius can be obtained. The accretion rate is $\dot{M}=\dot{M}_{\rm out}(R/R_{\rm out})^{\it s}$, where the possible wind is considered ($s$ is the wind parameter) and $\dot{M}_{\rm out}$ is the accretion rate at the outer radius, $R_{\rm out}$. In this work, we simply set $R_{\rm out}=10^4 R_{\rm g}$ ($R_{\rm g}$ is gravitational radius), and the wind parameter $s=0.4$ as constrained in Sgr A* and M87 \citep*[e.g.,][]{yuan03,feng17}. For other parameters in ADAF, we adopt the typical values that widely used in modelling LLAGNs and XRBs, where the viscosity parameter of $\alpha=0.3$, the ratio of gas to total pressure $\beta=0.5$, and the fraction of the turbulent dissipation that directly heats the electrons in the flow $\delta=0.1$ \citep*[see][for more details]{man00,liu13,yn14}. The radiation of synchrotron, bremsstrahlung, and Compton scattering are considered consistently in our calculations, where synchrotron photons are Compton upscattered by the hot electrons and produce the radiation in optical to X-ray waveband. At a sufficiently low $\dot{m}$, Comptonization become much weaker and the X-ray spectrum may be dominated by bremsstrahlung emission.  The $\dot{M}_{\rm out}$ is set as a free parameter.

Compared the accretion processes, the physics in the jet is much unclear \citep*[e.g., jet formation, jet acceleration,][]{cao16a,cao16b}. We assume that a small fraction of the accreting material was transferred into the jet (outflow rate $\dot{M}_{\rm jet}$). The shock will occur due to the collision of shells with different velocities in the outflow. We adopt the internal shock scenario that has been used to explain the broadband SEDs of XRBs, AGNs and afterglow of gamma-ray burst \citep*[e.g.,][]{pir99,yc05,wu07,nem14,xie16}. These shocks accelerate a fraction of the electrons, $\xi_{\rm e}$, into a power-law energy distribution with an index $p$. In this work, we fix $\xi_{\rm e}=0.01$ and allow the $p$ to be a free parameter that can be constrained from observations \citep*[e.g.,][]{yc05}. The energy density of accelerated electrons and amplified magnetic field are determined by two parameters, $\epsilon_{\rm e}$ and $\epsilon_{\rm B}$, which describe the fraction of the shock energy that goes into electrons and magnetic fields, respectively. Obviously, $\epsilon_{\rm e}$ and $\xi_{\rm e}$ are not independent. Only synchrotron emission is considered in calculation of the jet spectrum, where the synchrotron self-Compton in the jet is several orders of magnitude less than the synchrotron emission in X-ray band \citep*[see,][for more discussions]{wu07}. We set the size of jet as $\sim 10^5R_{\rm g}$, which is derived from the VLBA observations.  The jet viewing angle and jet velocity are unclear, we simply set $\theta_{\rm v}=20^o$ and $v_{jet}=0.6c$ in our calculations, where the sub-relativistic velocity are adopted as in most of nearby LLAGNs \citep[e.g.,][]{wu07,nem14,feng17}.  The free parameters are $\dot{M}_{\rm jet}$, $\epsilon_{\rm e}$, $\epsilon_{\rm B}$ and $p$ in the jet model.

%\begin{figure*}  \citep[][]{yn14}

%\includegraphics[width=8cm,height=6cm]{f1.eps}
%\caption{Multi-waveband spectrum for NGC 4993 and its ADAF-jet model results, where the dotted, dashed and solid lines represent the ADAF, jet and their total spectrum respectively.}
%\label{fig:2009p2013}
%\end{figure*}

 \begin{figure*}[]
 \centering
 \includegraphics[height=8cm,width=10cm]{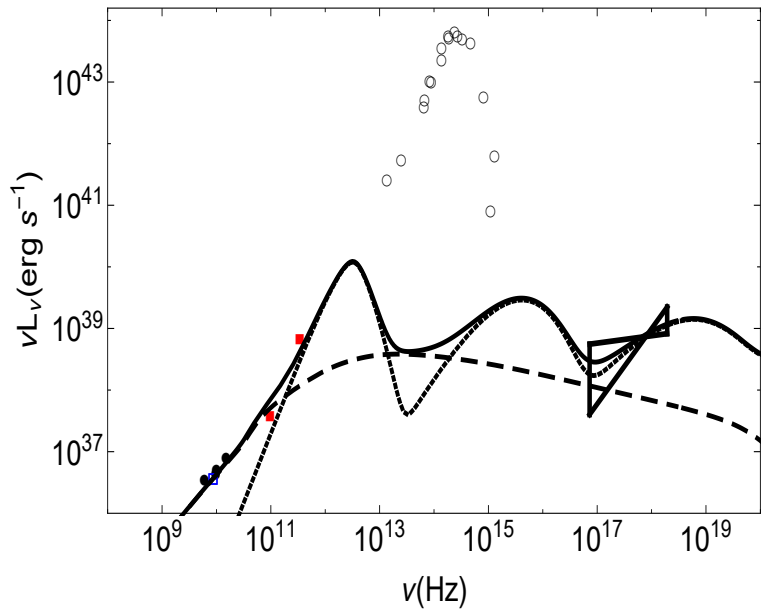}
 \caption{Multi-waveband spectrum for NGC 4993, where the solid circles represent the radio data observed by VLA, the open square represent the radio data observed by the VLBA, the solid squares represent the submm emission observed by AMLA, open circles represent the infrared-optical data observed by Spitzer/VLT/HST respectively. The bow tie is the X-ray observations. The dotted, dashed and solid lines represent the ADAF, jet and their total spectrum respectively. }
 \end{figure*}

\section{Result and Discussion} \label{sec:result}
   We find that the radio spectrum can not be simply described by a single power-law function ($F_{\nu}\propto \nu^{-k}$), where a break seem to exist at $\sim$ 100 GHz (see Figure 1). \textbf{For the radio observations at 6-100 GHz}, the power-law spectral index is $k\simeq0.14$, which is a slightly inverted spectrum. However, the spectral index is $k\simeq-1.3$ at 97.5-338.5 GHz that observed by ALMA, which is much steeper than that of low-frequency radio spectrum. This feature is also found in nearby LLAGNs \citep[e.g.,][]{an05,di05,di11,di16,pri16} and even in some Seyferts \citep[e.g.,][]{be15}. In particular, the submm bumps are very evident in two best studied nearby SMBHs, Sgr A* and M87 \citep*[e.g.,][]{an05,pri16}. This feature implies that they come from two different components. The X-ray spectrum is quite hard with $\Gamma=1.5\pm0.4$, which is consistent with other nearby LLAGNs. The mid-infrared to ultraviolet emission is much higher than those of other wavebands, which is measured with a quite large aperture (e.g., $\sim 1^{'}$) and may dominantly come from the host galaxies.

The central velocity dispersion is $\sigma_{*}\simeq 170\rm km\ s^{-1}$ for NGC 4993 \citep[][]{lev17}, the BH mass $M_{\rm BH}\simeq8\times10^{7}\msun$ can be obtained from the $M_{\rm BH}-\sigma_{*}$ relations \citep[][]{gu09}. We present our SED modeling for NGC 4993 in Figure 1. The steeper radio spectrum at $\sim100-300$GHz is roughly consistent with that prediction of ADAF, where the ADAF can also well explain the hard X-ray spectrum as observed by \textit{Chandra} (the dotted line). The accretion rate at $10^4 R_{g}$ is $\dot{M}_{\rm out}=5.3\times10^{-3}\dot{M}_{\rm Edd}$ and the accretion rate at the inner region of the accretion flow $\dot{M}(5R_{\rm g})=3.3\times10^{-4}\dot{M}_{\rm Edd}$. We note that most of radiation come from the accreting matter in the inner region of accretion flow near the BH horizon (e.g., within $10R_{\rm g}$), which is not sensitive to the accretion rate at outer boundary ($\dot{M}_{\rm out}$).  The low-frequency radio emission is much higher than the prediction of the ADAF and can be better explained by the jet model, where $\dot{M}_{\rm jet}=1\times10^{-6}\dot{M}_{\rm Edd}$, $\epsilon_e=0.06$,$\epsilon_B=0.02$ and $p=2.4$. It should be noted  that there is degeneracy in jet parameters (e.g., jet speed, outflow rate and magnetic field etc.), which will not affect our main conclusion on the origin of the radio emission. It is difficult for the jet model to explain the hard X-ray spectrum, where the jet emission can contribute at soft X-ray waveband in some level. Based on our modelling results, we find that the ratio of the mass-loss rate in the jet to the accretion rate estimated at $5R_{\rm g}$, $\dot{M}_{\rm jet}/\dot{M}(5R_{\rm g})\simeq4\times10^{-3}$. In other words, only a small fraction of 0.4\% of the mass that ultimately accreted by the BH is channeled into the jet. As expectation, the IR bump in the SED, measured with a larger aperture cannot fitted by the ADAF-jet model, which should dominantly come from the old stellar population in the host galaxy \citep*[e.g.,][]{nem14}. The detailed SED modeling of NGC 4993 allows us to put an independent constraints on the mass accretion rate into the BH and the jet mass-loss rate.

The origin of the multiwaveband emission in LLAGNs, and, in particularly in quiescent galaxies, is always in debate.  For the low-frequency radio emission in NGC4993, the unresolved emission is much higher than the prediction from star formation as constrained from the SED modeling of the host galaxies, where the radio spectrum is also much shallower than observed in star-forming galaxies \citep*[e.g.,][]{ba17}. The compact radio emission ($\sim mas$ scale or $<$ pc scale) detected by VLBA in the central region of NGC 4993 with a brightness temperature exceeding $10^6$K and $\sim$20\% of radio variability further suggest the presence of a LLAGN, and the radio emission should be not seriously contaminated by the host galaxy\citep*[e.g.,][]{and17,ax17}. The low-frequency radio emission (e.g., up to several tens GHz) is much higher than the prediction of the pure ADAF and the flat spectrum with $k\sim0.1$ is more consistent with the jet (e.g., Figure 1).

The high-resolution submm observations on the nearby LLAGNs are still very limited, where possible submm bump are found in two most well studied LLAGNs of Sgr A* and M87. The ALMA observations on NGC 4993 provide us the valuable information on the submm waveband, due to it is lucky as the host galaxy of the gravitational-wave event GW 170817A. The possible millimetre-wave excess (not the full submm bump) is also found in some of nearby LLAGNs and Seyferts \citep*[e.g.,][]{di05,di11,be15,di16}. \citet{di16} discussed the different physical origin of the high-frequency excess component based on a Seyfert galaxy NGC 985, where they found that the excess may originate from the free-free emission of the cloud in the broad emission line region, synchrotron jet free-free absorbed by a circumnuclear photoionized region and/or slef-absorbed nonthermal synchrotron from disk-corona, and rule out the possibilties of the dust emission at the Rayleigh-Jeans regime, compact jets under synchrotron self-absorption, or thermal synchrotron from ADAF. For NGC 4993, there is no far-infrared observations and we cannot constrain the possible contribution of millimeter emission from the dust even it is not important as found in NGC 985 \citep*[e.g.,][]{di16}. The free-free or bremsstrahlung emission is considered in our ADAF model, which may contribute at radio to X-ray waveband. However, our calculations show the radio and X-ray radiation from the free-free emission are much less than those from the Synchrotron and inverse-Compton radiation respectively for the same population of hot electrons in ADAF. Therefore, the free-free emission from the ADAF cannot explain the radio to millimieter emission in NGC 4993. The free-free emission from the possible broad-line region clouds should be also not important, since that there is no optical evidences for broad lines in NGC 4993. The steep millimeter emission should also not originate from the synchrotron self-abroption, where the spectral peak at $>300$ GHz is much larger than that of young radio galaxies (e.g., $\sim$ GHz) and it also cannot naturally explain the break at $\sim 100$ GHz as observed in radio spectrum \citep*[e.g.,][]{od98}. The mildly variability ($\sim$20-30\%) at 97.5 GHz within half-month further suggest that millimeter emission of NGC 4993 should mainly come from the BH activities, not from the host  \citep*[e.g.,][]{ax17}. \citet{mar08} proposed that the millimeter bump, including the other multi-waveband spectra, can be explained by the pure jet emission, where the millimeter emission was attributed to the quasi-thermal distribution of electrons in the nozzle at the base of jet. It should be noted that the synchrotron emission from the quasi-thermal electrons from the nozzle of jet should be more or less similar to that of ADAF. In our jet model, we didn't include this nozzle since that we naturally include the hot electrons in accretion flow. \citet{di16} found that the synchrotron emission from ADAF is insufficient for the millimeter excess in NGC 985, which may be caused by adoption of the old self-similar ADAF solution and an important fraction of gravitational energy in accretion flow can heat electrons directly in the updated ADAF model (e.g., $\delta=0.1-0.5$, see also Sect. 3).  It should be noted that the property of corona in bright AGNs should be more or less similar to that of ADAF (e.g., optically thin hot plasma with $T_{\rm e}\sim 10^9 K$), and, therefore, the millimeter excess in both LLAGNs and bright AGNs may be related to the hot plasma either in ADAF or corona \citep[e.g., ADAF/coronal outflow/wind, or not well collimated subrelativistic jet,][]{be15}. The multiwaveband correlation in variability may help to distinguish their possible origin.

It was suggested that the X-ray emission should be dominated by the ADAF and jet for the Eddington-scaled X-ray luminosity larger and less a critical value \citep*[e.g.,][]{yc05,wu07,yuan09}. Based on \citet{yc05}, the critical X-ray luminosity is $L_{2-10\rm keV,c}\sim 2\times10^{39}\rm \ergs$ for the BH mass of $\sim 10^8\msun$, and the observed X-ray luminosity of NGC 4993 is $L_{\rm 0.3-8keV}\simeq10^{39}\ergs$, which roughly corresponding to the critical luminosity. Our fitting result roughly support this scenario, where the X-ray emission is well consistent with the Comptonization emission from ADAF and just slightly higher than that predicted by the synchrotron emission from the jet. The ratio $L_{\rm R}/L_{X}\sim 5\times 10^{-3}$ in NGC 4993 suggest that it may be similar to other mildly radio-loud AGNs \citep*[e.g.,][]{be15}. The multiwaveband emission, including the submm waveband, may be fully dominated by the jet if the Eddington ratio is much less than this critical value, since that the ADAF emission ($L\propto \dot{M}^2$) decreases faster compared that of jet ($L\propto\dot{M}$) as the decrease of accretion rate. The jet emission may be more and more important in weak-activity BH sources. The radio emission, X-ray emission and BH mass of NGC 4993 follow the so-called the ''fundamental plane" of BH activities, which suggest that both the radio and X-ray emission should be dominated by the supermassive BH activities and was not seriously contaminated by the host galaxy \citep*[e.g.,][]{mer03}. %The simultaneous multiwaveband monitoring of this kind of LLAGNs will be a promising method to further explore their physical origins. 

\section{Conclusion} 
    Based on the most recent multi-waveband observations of the BH activities in NGC 4993, we find that the radio emission in millimeter waveband is much steeper than that of low-frequency radio band, where the break is also found in central BH of Sgr A* and M87, where the BH mass of NGC 4993 is more than one orders of magnitude larger than and smaller than that in Sgr A* and M87 respectively. We model the SEDs of NGC 4993 with a coupled ADAF jet model, and find that the steep millimeter radio spectrum and hard X-ray spectrum can be well explained by the ADAF model, while the low-frequency radio emission may come from the jet. The ratio of the outflow and inflow is $\sim0.4$\%, which suggests that only a small fraction of the accreting matter is channeled into jet if it exists. The submm bump may exist in SMBHs at different scales (e.g., $10^{6-10}\msun$) in low accretion regime, we suggest that the submm and deep X-ray observations will very helpful to help us to understand the central engines of nearby LLAGNs.

%\section{Conclusions} \label{sec:conclusion}

\acknowledgments
We thank the referee for many constructive comments that led to improvements in the paper. This work is supported by the NSFC (grants 11573009 and 11622324).


\end{document}